\documentclass[preprint]{IEEEtran}

\usepackage{cite}
\usepackage{amsmath,amssymb,amsfonts}
\usepackage{algorithmic}
\usepackage{graphicx}
\usepackage{textcomp}
\usepackage{xcolor}
\def\BibTeX{{\rm B\kern-.05em{\sc i\kern-.025em b}\kern-.08emT\kern-.1667em\lower.7ex\hbox{E}\kern-.125emX}}
\usepackage{enumitem}

\newlist{inlinelist}{enumerate*}{1}
\setlist*[inlinelist,1]{%
 label=(\roman*),
}

\renewcommand{\textbf}{\textit}
\NewCommandCopy{\svtt}{\texttt}
\renewcommand\texttt[1]{{\scalebox{.8}[1.0]{\svtt{#1}}}}

\begin{document}

\title{Blockchain Oracles for Real Estate Rental}

\author{\vspace{-1cm}}

\author{\IEEEauthorblockN{Nuno Braz$^1$ \ \ \ \
João Santos$^2$ \ \ \ \
Tiago Dias$^2$\ \ \ \
Miguel Correia$^1$} \\
\IEEEauthorblockA{\textit{INESC-ID, Instituto Superior Técnico, Universidade de Lisboa - Lisbon, Portugal} \\
\textit{Unlockit - Lisbon, Portugal} \\
\{nuno.braz, miguel.p.correia\}@tecnico.ulisboa.pt \{joao.santos, tiago.dias\}@unlockit.io
}}


\maketitle

\begin{abstract}
Blockchain technology has seen adoption across various industries and the real estate sector is no exception. The traditional property leasing process guarantees no trust between parties, uses insecure communication channels, and forces participants who are not familiar with the process to perform contracts. Blockchain technology emerges as a solution to simplify the traditional property leasing process. This work proposes the use of two blockchain oracles to handle, respectively, maintenance issues and automate rent payments in the context of property rental. These two components are introduced in a blockchain-based property rental platform.
\end{abstract}

\section{Introduction} 

Blockchain technology is gaining popularity and is being used in various industries \cite{nakamoto2008bitcoin,peck2017blockchains,yaga2019blockchain,dinh2018untangling}. This work focuses on its application to \emph{real estate}, specifically to the \emph{rental of properties} \cite{cuttell2017blockchain,qi2019blockchain,xue2021housing,shanker2019use,sharmaimplementation,santos2024brains}. The traditional property rental process has evolved with technological advancements, yet there remains room for improvement. Opportunities start with the complexity of the property rental market, which involves not only state regulations but also considerations related to lease types. Moreover, documentation is exchanged through multiple insecure channels, such as e-mail and messaging applications. This violates privacy laws such as the European General Data Protection Regulation \cite{gdpr2016general}. The present practice for creating rental agreements forces participants who are unfamiliar with the process to carry out different types of contracts. More critically, the different stakeholders do not trust each other. This lack of trust in the property rental process drives some homeowners away from the market and forces others to increase rent prices to cover potential losses. This is a particularly important issue due to the housing crisis that has intensified in recent years.

Blockchain can provide consistency and additional trust in the property rental process. The use of smart contracts to enforce lease agreement conditions reduces the need for the entities involved to trust each other and to know how to force the agreed clauses. Permissioned ledgers allow users to control what data they share and with who they share it with. Furthermore, these ledgers allow the imposition of entry barriers for new users, supporting Know Your Customer (KYC) processes \cite{gill2004preventing}, thus enhancing security. Transactions are facilitated by the creation of smart rental contracts, i.e., the implementation of a rental contract as a smart contract. Smart contracts are immutable code that runs on the blockchain. These smart rental contracts automatically execute contractual agreements, this automation enhances trust and increases process efficiency. Due to its immutable nature, information stored in the blockchain cannot be altered, providing a trustworthy and unchangeable record. The blockchain creates a trusting environment between the tenant and the landlord, mitigating fraud opportunities. 

Decentralization on blockchains is achieved through consensus mechanisms. These mechanisms follow deterministic rules to ensure that all nodes reach the same conclusion about the state of the blockchain. For example, given the same set of transactions and state, every node should end up with the same final state after the process. This restricts the execution of smart contracts to the data already stored on-chain. However, most of the capabilities of smart contracts are lost without access to external data. 

Blockchain \emph{oracles} are services that enable smart contracts to interact with real-world data. Many types of oracle exist with different characteristics and functionalities \cite{beniiche2020study,muhlberger2020foundational,mammadzada2019blockchain}. They can be applied in many different systems, operating differently depending on what they were designed for. Implementing oracles poses significant challenges as they can be seen as potential centralized points of failure in a decentralized environment or introduce concerns related to security and trust. 

\emph{The paper proposes two oracles} to make the rental process more trustworthy, specifically during the lease period by addressing two key challenges: delayed rent payments and determining responsibility for property issues that arise during the lease; in the document, these are referred to as Maintenance Issues (MIs).  These problems can include scenarios such as a broken chair, a power failure, or a broken window due to bad weather. Using the immutability of a blockchain, it is possible to store the details of all the issues and interactions between the parties. 
To assess the benefits of these two oracles, a \emph{blockchain-based property rental platform} developed using the Daml ledger model \cite{DAML} is also presented. The platform integrates the two oracles: Automated Rent Collection Oracle (ARC-Oracle) to streamline rent payments; and Maintenance Issues Oracle (MI-Oracle) to manage and document property-related problems. The former is fully automated, but the latter obtains its data from humans, which we call arbitrators. 
We evaluate the oracles in terms of performance and security.

\section{Background}

%
Daml is a functional programming language designed
to implement smart contracts \cite{DAML}. The Daml ledger model is a virtual permissioned ledger that allows workflows among different parties. These workflows are written in Daml and executed as smart contracts. A party can be a legal entity, a physical person, or just one of many accounts for an entity or person. The Daml interpreter deterministically converts a Daml expression into a transaction, and the Canton protocol then ensures that the transaction is performed securely and atomically. The Canton protocol allows one to create distributed ledgers by handling synchronization, security, and privacy by itself, allowing developers to focus solely on business logic \cite{Canton}. Authentication and data transport are handled through their synchronization domains (sync domains). These domains allow communication between participant nodes, by routing messages to provide a consistent transaction order and confirming transaction commits. The sync domains can be built with different levels of decentralization.

Oracles are services that connect the outside world to blockchain smart contracts. They can be standalone or provide a user interface (UI). 
Oracles can be classified according to the source of information, the direction of information (from the inside of the blockchain to the outside or from the outside of the blockchain to the inside), and the level of trust \cite{beniiche2020study}. Designing an oracle possesses several challenges, namely the dilemma between efficiency and decentralization. Centralized oracles consist of a single node managed by a single operator. Distributed but not decentralized oracles are made up of multiple nodes spread across various locations but rely on a single infrastructure provider and operator. Decentralized oracles are made up of multiple nodes distributed in different locations, with different infrastructure providers and operators. Chainlink takes a step further with a network of decentralized oracles \cite{ChainlinkEducation}. The more decentralized the oracle, the more reliable and secure it is. These oracles achieve truthfulness by using consensus algorithms. However, the consensus algorithms reduce their response time. 



\section{Related Work}

When talking about blockchain-based rental systems, there are a few start-up companies working in the area \cite{RentPeacefully}, \cite{SMARTRealty},\cite{RentibleWhitepaper}. There are also academic works that present such systems \cite{qi2019blockchain,cuttell2017blockchain,sharmaimplementation,xue2021housing,shanker2019use}, but do not consider oracles to solve the problems this paper handles. Closer to our work, \cite{cuttell2017blockchain} presents the \textit{Arbitrator} entity, a human source of information whose purpose is to solve disputes between tenants and landlords, and \cite{sharmaimplementation} implemented a reputation system.

Regarding user verification, Rentible \cite{RentibleWhitepaper} has implemented a KYC protocol associated with the user's wallet. Since this protocol ensures authentication with the user itself and not with an account, it increases trust between all users of the platform and reduces the risk of fraud because the same person creating several accounts is not possible. 
Qi-Long et al. \cite{qi2019blockchain} and Sharma et al. \cite{sharmaimplementation} use IPFS to store the listing information. Being a decentralized protocol ensures that the data used in the system has not been tampered with. 
Sharma et al. \cite{sharmaimplementation} reputation system adds credibility to the rental system because only users with a confirmed tenant-landlord relationship can do reviews and this relationship is assured by the immutability of the blockchain. 

Some works propose manual payments and payments triggered by users \cite{Proenca:2023,cuttell2017blockchain}. Cuttell et al. \cite{cuttell2017blockchain} present an automation mechanism that is more secure since tenants do not have to trust their password with a third party and it addresses the problem of the account not having enough funds before payment. Automatic payments also solve the problem of delayed payments but must be triggered by an external entity to the blockchain. 
Xue et al. \cite{xue2021housing} include government entities such as the Certification Department in the system. This adds trust to the system because the tenants are who they claim to be and the property is legitimately owned by the landlord. 

In a blockchain system, it is not possible to schedule transactions without an off-chain entity, such as a human (manual input) or a centralized script. However, these two solutions induce centralization risks, efficiency bottlenecks, and security risks. 
Two proposed solutions for this problem are Aion's system and Chainlink's Upkeep contracts \cite{AionDocumentation}, \cite{ChainlinkAutomationDocs}.

\section{The Platform and the Oracles}

The paper proposes two oracles, but also the platform used to assess them. The platform uses the Canton Protocol, a permissioned blockchain, to control the actions each entity can perform on the platform and to allow each user to maintain the privacy of its data. In the Canton Protocol, each node has one or more parties connected to it. The nodes connect to synchronization domains (sync domains). These domains handle authentication and data transport. In the sync domain, the sequencer is responsible for ordering messages and delivering them to the participants for validation, as well as timestamping the transactions. The mediator receives the responses from the participants and aggregates them into a single verdict for the entire confirmation request, and it is used to hide the participant's entities from each other.

\subsection{Platform and Oracles Overview}

This platform supports the workflow of the creation of a Lease Agreement between \textit{Tenants} and \textit{Landlords}, plus the additional functionalities provided by the oracles. It is in line with the privacy and authorization required for these workflows. 
The ARC-Oracle functions as a daily trigger to process rent payments. The payment of the rent is considered to be itself a smart contract, called an IOU (I Owe You), that acknowledges the existence of a debt. The MI-Oracle is used to solve disputes when MIs occur during the lease period. A maintenance issue (MI) is associated with a lease agreement. It has a starting date, a duration, and a repair cost. When resolved, it will have a percentage of responsibility attributed to the \textit{Landlord} and the \textit{Tenant}. The resolution is represented by a smart contract with the MI details and the percentage of responsibility.

\subsubsection{Participants}

\begin{figure}[t]
  \centering
  \includegraphics[width=0.5\linewidth]{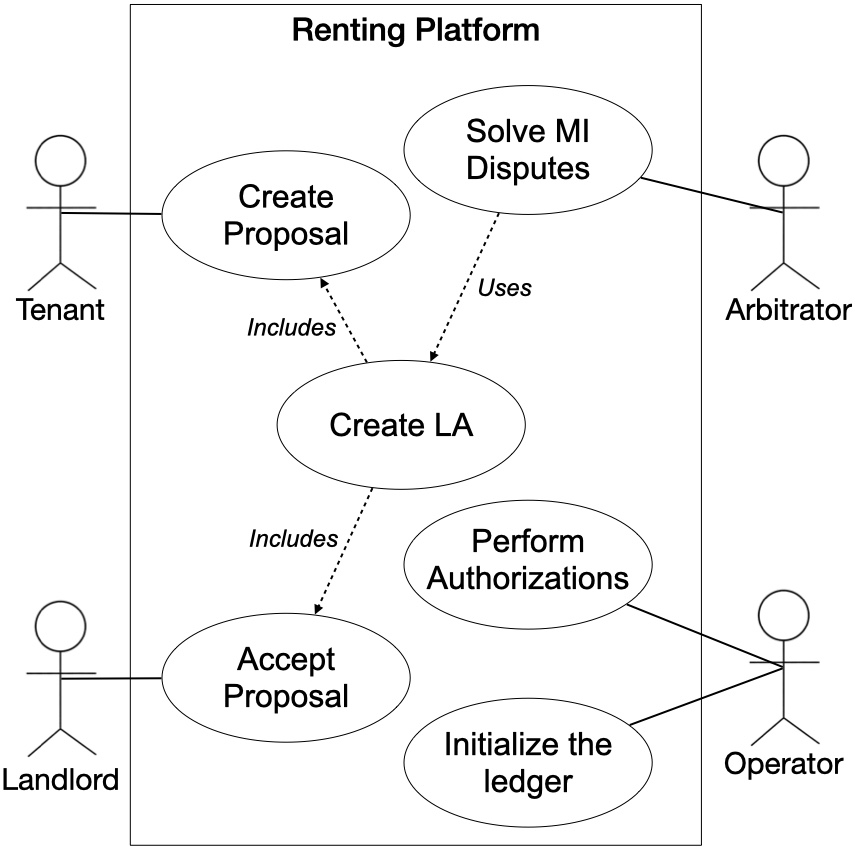}
  \caption{General use case diagram}
  \label{fig:generalUseCaseDiag}
\end{figure}

The users of this platform are the \textit{Tenant}, who rents the property, and the \textit{Landlord}, who grants the right to use the property in exchange for the rent. \textit{Tenants} and \textit{Landlords} are the \textit{Users}. In addition, there are two more human entities on the platform, the \textit{Operator} and the \textit{Arbitrator}. The former is considered a source of trust: he initializes the ledger and is a stakeholder in the contracts that rely on outer entities. The latter has the role of providing a report of any MI that can occur during the lease period (see Fig.\ \ref{fig:generalUseCaseDiag}).

\subsubsection{Lease Agreement Creation}

For simplicity, on this platform, the Lease Agreement (LA) creation process begins with the \textit{Tenant} submitting a Proposal. This is done through his front-end interface, User UI, specifying the House to rent and the Lease Terms. The Lease Terms contain the rent, beginning date, payment dates, and the number of arbitrators to be called when an MI occurs. The \textit{Landlord} can then accept or reject the Proposal. Before the \textit{Landlord} makes a decision, the \textit{Tenant} has the option to withdraw the Proposal. If the Proposal is rejected or withdrawn, the \textit{Tenant} can submit a new Proposal with different Lease Terms. An LA Request contract will be generated if the \textit{Landlord} accepts the Proposal. The LA contract will be created with the information from the LA Request once this request is validated by the \textit{Operator}.

\subsection{System Architecture}

\begin{figure}[t]
  \centering
  \includegraphics[width=\linewidth]{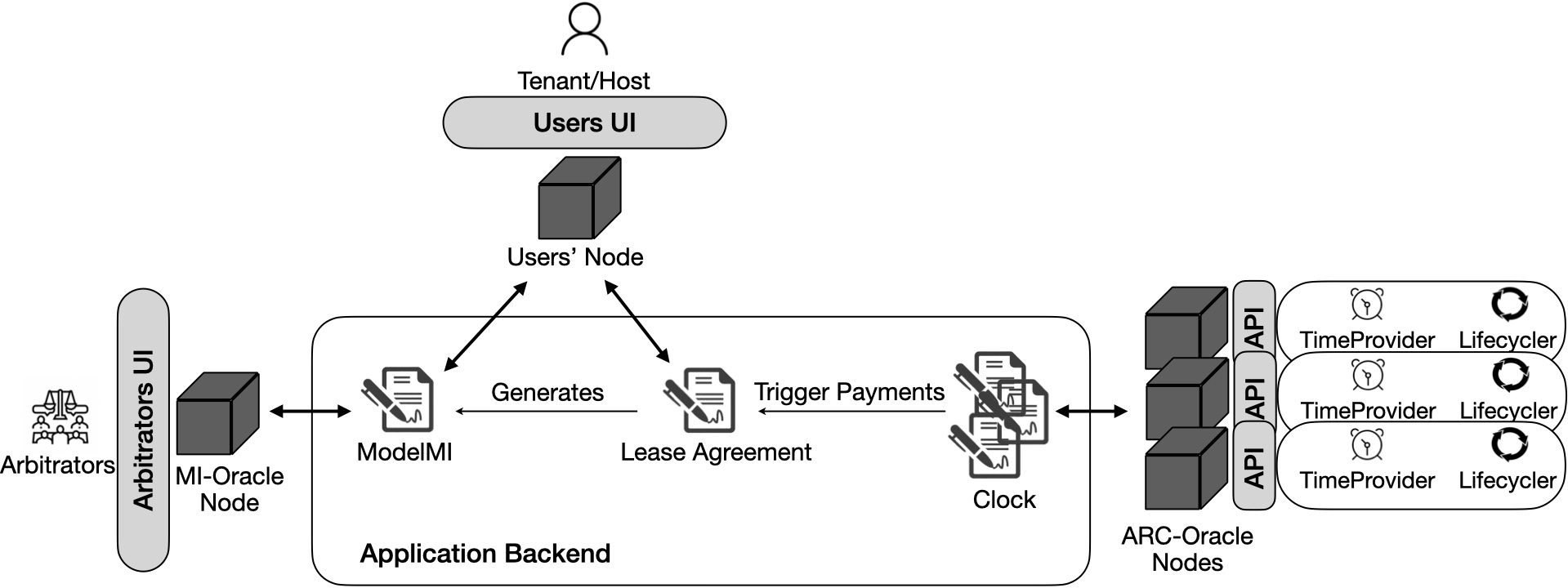}
  \caption{General system architecture}
  \label{fig:sysArchitecture}
\end{figure}

The architecture is represented in Fig.\ \ref{fig:sysArchitecture}. It consists of an application back-end to manage smart contracts, two UIs, one for the users and the other for the \textit{arbitrators}, and an API to connect the schedulers to the oracle node. 
\textit{Users} interact with the blockchain-based platform via a User UI, which communicates with the \textit{Users' node} to process actions such as managing lease agreements or reporting MIs. The node interacts exclusively with the MI-Module and LA-Module.

The MI-Oracle has a similar design. The \textit{Arbitrators} interact with the Arbitrator UI, which communicates with the \textit{MI-Oracle node} that interacts exclusively with the MI-Module. 
Each \textit{User} and \textit{Arbitrator} is represented as an individual party and must trust the \textit{Users' node} and \textit{MI-Oracle node} respectively. The users
do not need to trust other parties, the \textit{Arbitrators} are assumed to be trusted parties. Each of the nodes can be replicated to scale the application. However, before joining, each node must be approved by the sync domain.

The ARC-Oracle consists of a scheduler implemented in the ARC-Oracle node. The node interacts only with the ARC-Module.
In the figure, \textit{TimeProvider} and \textit{Lifecycler} belong to the ARC-Oracle node. However, from the perspective of the smart contract's back-end, each party has a different function. The \textit{TimeProvider} updates the platform date clock. The \textit{Lifecycler} initiates the rent collection workflows. Optionally, a node instance can be configured where both \textit{TimeProvider} and \textit{Lifecycler} functions are executed by the same party.

\subsection{Automatic Rent Collection Oracle (ARC-Oracle)}

\begin{figure}[t]
  \centering
  \includegraphics[width=1\linewidth]{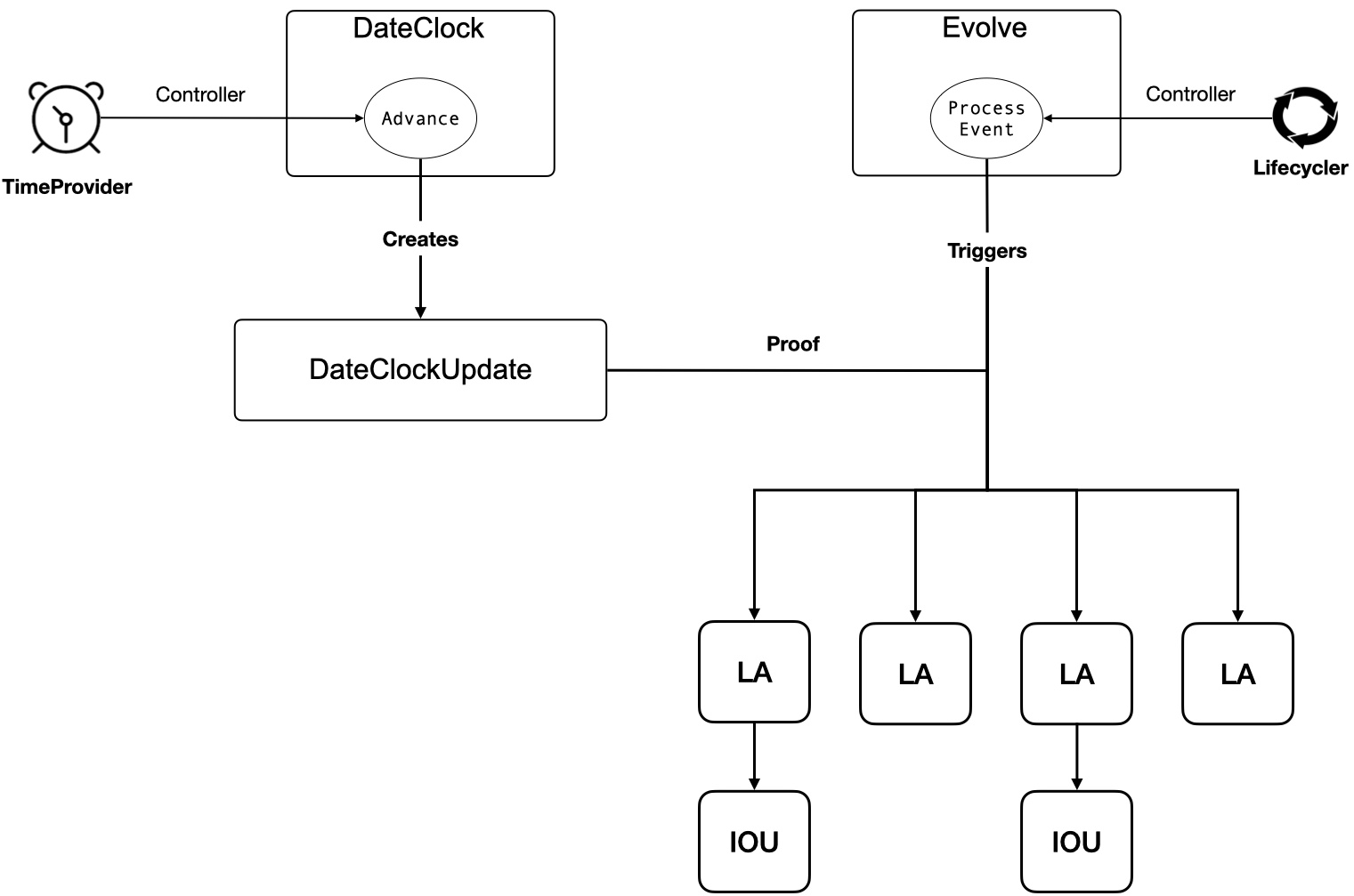}
 \caption{ARC-Oracle architecture}
  \label{fig:rentCollection}
\end{figure}

The ARC-Oracle can be classified as a push-based inbound oracle. It does not have a data source because it functions purely as a trigger; there is no need to retrieve data from external sources. The oracle solution is composed of a set of nodes (\textit{ARC-Oracle nodes}) and the ARC-Module. The architecture of the ARC-Module is shown in Fig.\ \ref{fig:rentCollection}.

Two transactions are executed daily to collect the rent payments. The first is the \texttt{DateClock} smart contract update by the \textit{TimeProvider}. The \texttt{DateClockUpdate} smart contract is created within this transaction. 
The second is to process the date update. The \textit{Lifecycler} initiates the collection of rent in the \texttt{Evolve} contract using \texttt{DateClockUpdate} as proof of the passing of the day. The \texttt{Evolve} contract, which stores a set of Lease Agreements (LA) on the platform, identifies lease agreements with overdue payments and generates an \texttt{IOU} contract for each, representing the \textit{Tenant}'s debt to the \textit{Landlord} with the value of the rent. 

In the \texttt{DateClock} update, the new date is set to the value of the transaction timestamp. 
In Daml each transaction is associated with two timestamps: Ledger Time and Record Time. Ledger Time is set by the submitting participant and the Record Time is assigned by the sync domain when the transaction is recorded. The Record Time is acknowledged as accurate. The model guarantees that the Record Time and Ledger Time deviate by no more than the skew value. The Ledger Time can be obtained by calling the \texttt{getTime} function \cite{damltime}
The correctness of this value depends on the sync domain and the submitting party. Tampering this value would require collusion between these two entities. 

The \texttt{DateClockUpdate} is created to prevent making the \texttt{DateClock} public, as this would allow any user to try to submit a transaction to advance the clock. Although tampering with the date would still require the agreement of the synchronization domain, it could expose the system to Denial of Service attacks and block the \textit{TimeProvider} from advancing the time. The \texttt{DateClockUpdate} provides the current date to the whole system. By referencing this contract instead of calling the \texttt{getTime} function, it ensures that if two transactions that require the date occur back-to-back at midnight, they will share the date, or only the second transaction will have the date of the following day. The first transaction can never have the date after the second transaction. This cannot be assured when calling the \texttt{getTime} function, since there is a skew value between the record time and the ledger time \cite{damltime}. This mitigates races when two transactions happen back-to-back at midnight.


\subsection{Maintenance Issues Oracle (MI-Oracle)}

\begin{figure}[t]
  \centering
  \includegraphics[width=0.9\linewidth]{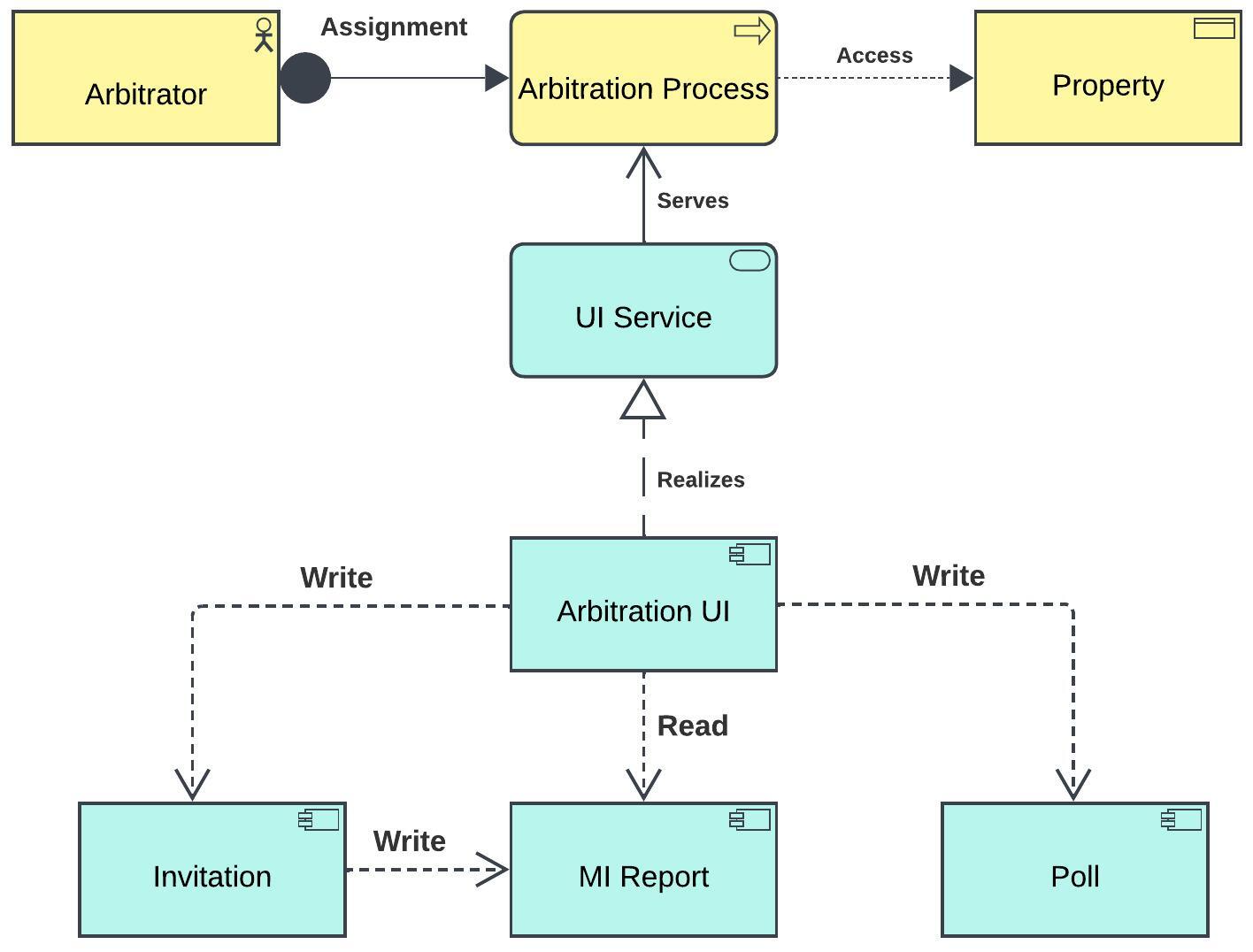}
 \caption{MI-Oracle Archimate diagram}
  \label{fig:miarchi}
\end{figure}

The MI-Oracle is a Human Oracle. It can be classified as a pull-based inbound oracle. The MI-Oracle architecture is simple since it retrieves data from a human data source. The Arbitrator UI manipulates three smart contracts: Invitation, MI Report, and Poll. As shown in Fig.\ \ref{fig:miarchi}, the UI is used by the \textit{Arbitrator}. For each MI, one of the \textit{Arbitrators} will be a \textit{Visitor}. Its role is to physically visit the property to evaluate the reported issue and to create a poll. This poll allows the rest of the \textit{Arbitrators} to vote on the level of responsibility assigned to each stakeholder. There is no specific process for selecting the \textit{Visitor}. However, its identity is recorded on-chain.

A MI Report can be created by the \textit{tenant} or the \textit{landlord} if any property problem arises during the lease period. The report contains the starting date and the description of the issue. 
In the platform there is the option for MI issues to be resolved by mediation; this is out of the scope of the MI-Oracle. The resolution by mediation is a more convenient option for both the \textit{Tenant} and the \textit{Landlord} since it does not bear the additional costs of allocating \textit{Arbitrators} in this process, smart contracts are used as the mediators for the resolution. The resolution by mediation is not part of the formal arbitration process; rather, it serves as an additional step in the MI resolution process or as an alternative to it.

The Arbitration process is more complex and involves two phases, the invitation and the polling. Triggering this process is a unilateral decision by the \textit{Tenant} and \textit{Landlord}, but both parties can perform the same actions throughout the procedure. 

\begin{figure}[t] 
  \centering
  \includegraphics[width=\linewidth]{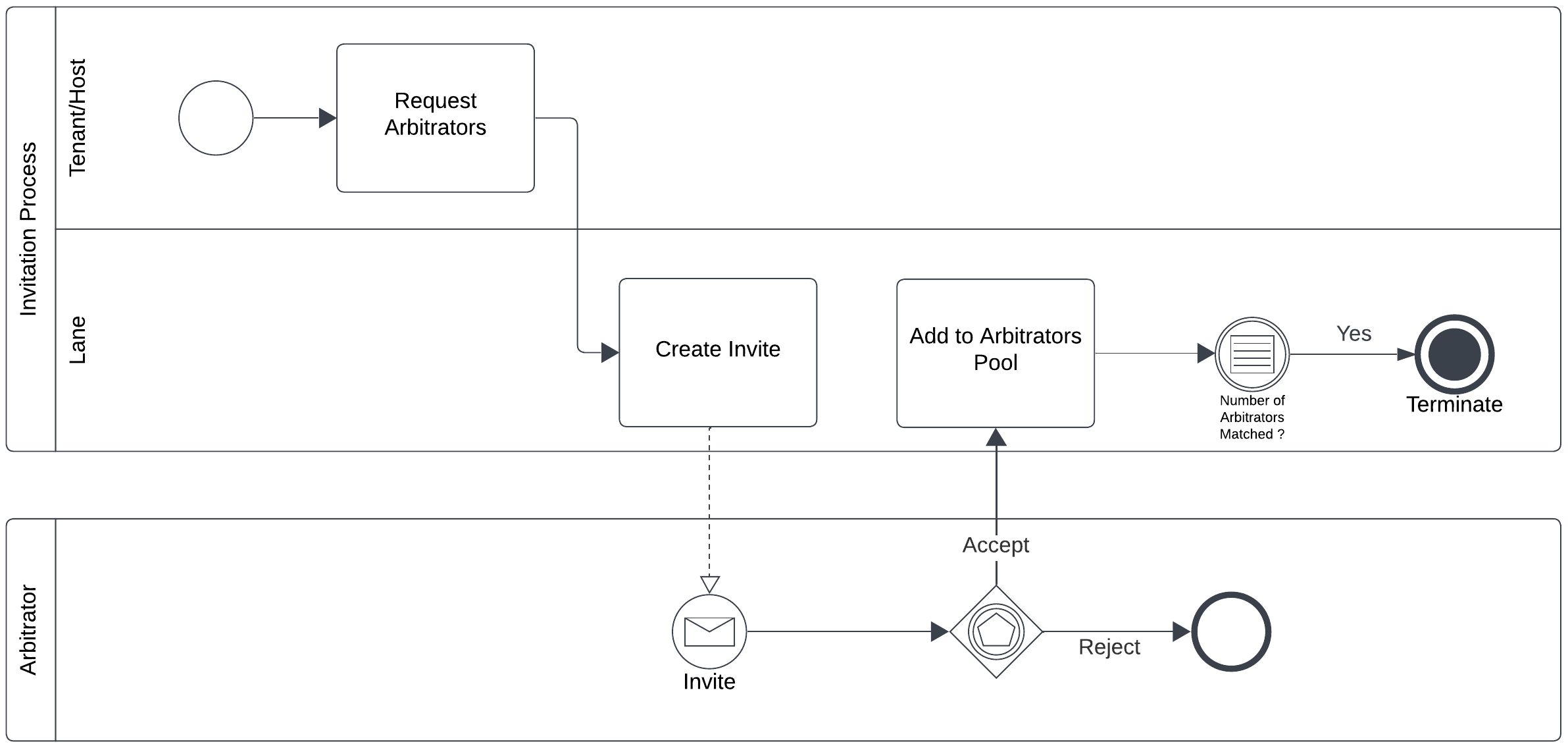}
 \caption{BPMN diagram of the \textit{Arbitrators} invitation process}
  \label{fig:InvitationBpmn}
\end{figure}

The invitation process is shown in the
diagram of Fig.\ \ref{fig:InvitationBpmn}. The invitation is sent to all available arbitrators. 
This invitation remains active until the required number of arbitrators is reached. It is important to state that only legitimate \textit{Arbitrator} entities are invited. Having an off-chain list of the legitimate \textit{Arbitrator} entities on the user client is not a safe approach since it could easily be tampered with and there would be no point in using a DLT. To address this, a publicly accessible smart contract \texttt{AvailableArbitrators} maintains a list of all available \textit{Arbitrator} parties. Only the \texttt{Operator} entity can add new \textit{Arbitrators} to the smart contract list.

\begin{figure}[t]
  \centering
  \includegraphics[width=0.8\linewidth]{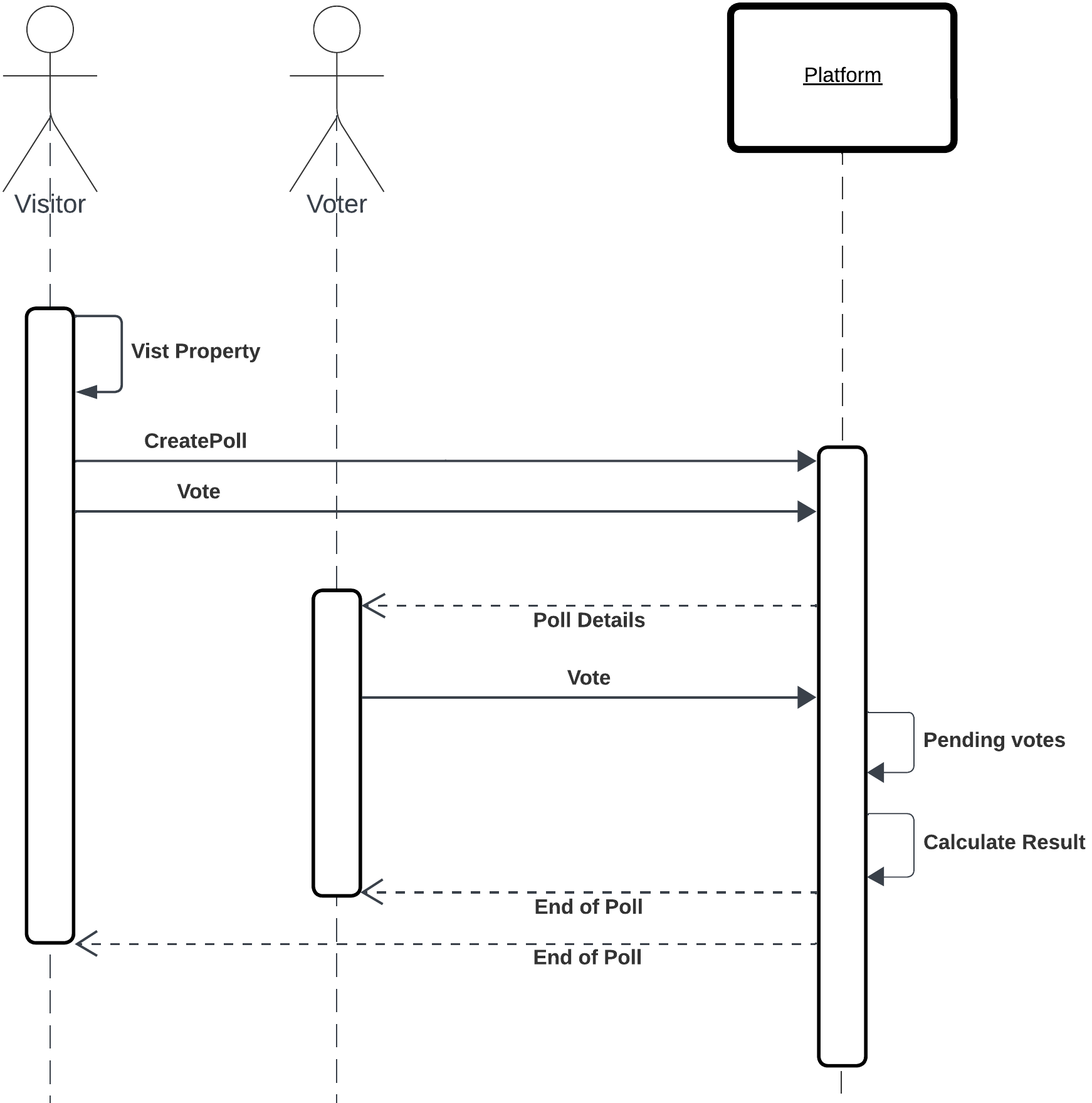}
 \caption{\textit{Arbitrators} polling process}
  \label{fig:pollingSeq}
\end{figure}

The polling process shown in Fig.\ \ref{fig:pollingSeq} is pretty straightforward. After the \textit{Visitor} evaluates the issue in the property, it creates the \texttt{Poll} smart contract that includes all the relevant details for the other \textit{Arbitrators} to provide an accurate vote.

A decision on the \( N \) arbitrators to attribute to an MI will be required to assess a trade-off between cost and performance. A higher value of \( N \) will give us a better assessment, yet it comes at the expense of increased resource utilization; therefore, the process fees vary depending on the value of \( N \). In our implementation the \textit{Tenant} and \textit{Landlord} agree on the value of \( N \) in the lease terms. Other options are: fix a \( N \) balancing cost and performance, for each MI during the lease period the \textit{Tenant} and \textit{Landlord} agree on the \( N \) to use which provides a better balance assessment. 


The MI protocol requires the \textit{Arbitrators} to be trusted. Starting by the \textit{Visitor}, to corrupt the system it would be enough to submit a wrong evaluation of the MI details. 
Next, we explain
how to achieve a fair MI resolution even if some of the \textit{Arbitrators} are corrupt. 

In this situation, it can be beneficial to create a reputation system to remove corrupt \textit{Arbitrators} from the system and a reward system to encourage the correct behavior of each \textit{Arbitrator}. The \textit{Arbitrators} that provide feedback that deviates from the other feedback must be penalized in the reputation system. Using a reward system and a higher number of \textit{Arbitrators}, corrupt the system by bribing the \textit{Arbitrators} would be more costly. Since one would have to bribe more entities and at a higher price. Attributing the rewards can be challenging, it must be more profitable for the \textit{Arbitrator} to submit an accurate vote than to follow an incentive to corrupt the system. For example, a bribe could serve as an incentive for corruption. A party may find it beneficial to bribe the \textit{Arbitrator} with an amount lower than the cost of fixing the problem. However, it is not economically viable for the platform or its users to reward each \textit{Arbitrator} that amount for its correct behavior. One way to address this is to keep the identity of each Arbitrator confidential from the others, as well as from the \textit{Tenant} and \textit{Landlord}. Smaller rewards could be offered to the \textit{Arbitrators}, for example, assigning them to more profitable disputes. In a permissioned blockchain, this is possible by creating an instance of the MI smart contract to each \textit{Arbitrator}.
The votes must remain confidential until all \textit{Arbitrators} have voted. There are two ways to achieve this. Using a commit and reveal scheme. Where the \textit{Arbitrators} cast their vote hashed with a password that can only be revealed after all votes have been submitted. Alternatively, by leveraging Daml's privacy guarantees, it is possible to create smart contracts with the votes visible only to the voter and a trusted party, such as the \textit{Operator}, or the \textit{Tenant} and \textit{Landlord}. This party would then compute the result.
The outcome of all votes must not be vulnerable to manipulation by a single voter. For example, if the results are determined by calculating the mean of all votes, submitting an extreme value above or below the mean could easily shift the result.
Keeping the identity of the \textit{Visitor} confidential from the \textit{Tenant} and \textit{Landlord} is also a challenge. Having more than one \textit{Visitor} also reduces the risk of bribery. 


\section{Platform Implementation}

This section provides a description of the implementation of a prototype to demonstrate the use of oracles.
The Daml ledger guarantees authorization and privacy, i.e., that all actions that affect a party were previously authorized by him and that if a party is not involved in an action it does not know it happened. 

\subsection{Back-end Implementation}

The back-end is implemented using the Daml language, and to fully understand the implementation decisions is important to understand the Daml ledger model. In the Daml ledger model it is not possible to prevent contracts from being created directly, which means that any party can create any contract. 

In Daml, contracts are created using templates. Contracts contain data, referred to as the \textit{create arguments}, these arguments have a name and a type. Unlike Java classes, a contract cannot change its \textit{create arguments}, a new contract must be created. We refer to the \textit{create arguments} as \textit{attributes} because most time we can think about a contract as a class. Each contract must have one or more signatories, parties whose authority is required to create the contract or archive it. Additionally, a contract can have observers, parties who can see the contract and all actions performed on it. When a new instance of a contract is created, a contract Id is attributed to it. 

If we think about the contract templates as classes and contracts as objects, choices would be the methods. They are predefined actions that can be exercised in the contract. They allow the signatories to delegate the right to perform a data transformation in a contract. Daml choices have controllers, which are the parties that have the right to exercise them.

\subsubsection{Lease Agreement Creation}

The implementation of a Lease Agreement Proposal follows a Propose and Accept Pattern to ensure both parties sign the Lease Terms proposed. It functions as follows, the \textit{Tenant} creates a \texttt{Proposal} contract that includes the \texttt{LeaseTerms}, the \texttt{House} they intend to rent, and the \textit{Operator} party. Within this contract, the \textit{Landlord} can exercise the \texttt{Decline} choice and the \textit{Tenant} can exercise the \texttt{Withdraw} choice; both options will archive the contract. The \textit{Landlord} can also exercise the \texttt{Accept} choice, which archives the contract and creates a \texttt{LACreationRequest} contract. In the \texttt{LACreationRequest} contract is required the \textit{Operator}'s permission to create the \texttt{LeaseAgreement} contract.

\subsubsection{ARC-Oracle}

The automatic rent collection back-end was designed with scalability in mind while also addressing potential threats from malicious actors. This process can be divided into two parts: updating the \texttt{DateClock} contract and \textit{lifecycling} the rent payments. In this context, the term \textit{lifecycle} means to verify in all lease agreements if there are due payments and, if so, to process them. 

In the smart contracts implementation, the \textit{TimeProvider} and \textit{Lifecycler} considered different parties.

The \texttt{DateClock} contract has the following attributes. The \texttt{providers}, a set of parties that are authorized to advance the date in the contract. This is needed for when there is more than one node communicating with the same ARC-Module, to increase fault tolerance. The \texttt{creator}, which is the last party to update the date. The \texttt{waitingAccept} that is used to add providers to the contract. This is done by the \textit{Operator}.

To update the date, one of the providers exercises the \texttt{Advance} choice in the contract. This choice calls the \texttt{getTime} function that returns the value of the transaction timestamp. The current date is obtained from this value and an offset is calculated from this date value and the date on the \texttt{DateClock} contract. If this offset is larger than 0 then a new \texttt{DateClock} and \texttt{DateClockUpdate} are created, and the old ones are archived. .

The \texttt{Evolve} contract contains a set of \texttt{laKeys}. These are the keys to every lease agreement associated with the \texttt{Evolve} contract. The signatory of this contract is the \textit{Operator}.

There are two choices in this contract. The \texttt{AddLA} choice adds a laKey to the \texttt{lakeys} set in the \texttt{Evolve} contract. It can only be exercised by the \textit{Operator}. And the \texttt{ProcessEvent} choice, is exercised by either the \textit{Lifecycler}. This choice iterates through all \texttt{laKeys} and exercises the \texttt{ProcessPayment} choice on each \texttt{LeaseAgereement} contract associated with the key. The \texttt{ProcessPayment} choice requires the \texttt{DateClockUpdate} contract as proof of the current date and checks which payment dates are earlier than the date specified in the \texttt{DateClockUpdate}, for these dates it creates an \texttt{IOU} with the \textit{Landlord} as owner and the \textit{Tenant} as debtor.   

\subsubsection{MI-Oracle}

The MI process is triggered by calling the \texttt{CreateMI} choice in the \texttt{LeaseAgreement} contract with the following parameters: \texttt{house}, \texttt{description}, and \texttt{startingDate}. This choice creates the \texttt{MIReport} contract below.

\begin{footnotesize}
\begin{verbatim}
MIReport
    tenant : Party
    landlord.: Party
    laKey : LAkey
    miDetails: MIdetails 
    arbitrators : Parties
    activeInvitation : Bool
    
    signatory tenant, landlord
    observer arbitrators
\end{verbatim} 
\end{footnotesize}

The \texttt{MIReport} is initialized with the \texttt{arbitrators} attribute as an empty set.

The resolution by mediation is triggered by the \texttt{SubmitAssessment} choice which requires the following parameters: the creator (either the \textit{Tenant} or \textit{Landlord}), the assessment details, which includes the responsibility of each party and the cost of repairs, and the contract Id of the associated \texttt{MiReport} contract. This contract Id is used to fetch the real number of arbitrators agreed in the lease terms. This choice creates the \texttt{MediationAssessment} with the creator as a signatory and the counterpart as an observer. To accept this assessment, the counterpart must sign this contract by exercising the \texttt{AcceptAssessment} choice in the \texttt{MediationAssessment} contract. This choice archives the contract and creates the \texttt{MIresultMediation} contract. To reject the assessment, the counterpart exercises the \texttt{Reject} choice that archives the contract.

In the resolution by arbitration the first step is to assign \textit{Arbitrators} to the \texttt{MIReport} contract.
Only legitimate \textit{Arbitrator} entities must be invited. This is achieved with the \texttt{AvailableArbitrators} and \texttt{AvailableArbitratorsRequest} contracts below.

\begin{footnotesize}
\begin{verbatim}
AvailableArbitrators
    operator : Party
    arbitrators : Parties
    observers : Parties
    
    signatory operator
    observer observers
\end{verbatim} 
\end{footnotesize}

\begin{footnotesize}
\begin{verbatim}
AvailableArbitratorsRequest
    public : Party
    requester : Party
    
    signatory requester
    observer public
\end{verbatim} 
\end{footnotesize}

To create the \texttt{InviteArbitrators} contract where the \textit{Arbitrators} can accept the invitation. One of the parties, \textit{Tenant} or \textit{Landlord} exercises the \texttt{InvokeArbitrators} choice in the \texttt{LeaseAgreement} contract. With an \texttt{AvailableArbitrators} contract Id and \texttt{MIReport} contract Id as parameters. 

It is impractical to include all platform users in the observer set of the \texttt{AvailableArbitrators} attribute. Moreover, some of these users may never interact with this contract. The following describes a more scalable approach.
The \texttt{observers} attribute of the \texttt{AvailableArbitrators} is initialized with the public party. All users can read and act in the name of this party. However, this party cannot be an observer on the \texttt{LeaseAgreement} for privacy reasons. Therefore, we have to create an instance of the \texttt{AvailableArbitrators} contract where only the party that will invite the \textit{Arbitrators} is an observer. For this, we use \texttt{AvailableArbitratorsRequest} that any party can create. By providing the \texttt{AvailableArbitratorsRequest} contract Id as a parameter, anyone can exercise the \texttt{AddObserver} choice. This creates a new instance of the \texttt{AvailableArbitrators} contract with the requester as an observer. 

The \texttt{InvokeArbitrators} choice starts by verifying that there is no Invitation Process currently active, through the \texttt{activeInvitation} flag. If not, it sets it to \texttt{true}. Then it fetches the attributes of the \texttt{MIReport} contract and archives it. The instance of the  \texttt{AvailableArbitrators} contract created previously is fetched to retrieve the \texttt{arbitrators} attribute and then archived. The \textit{Operator} is the only signatory of the \texttt{AvailableArbitrators} contract and, therefore, is the only entity with the right to archive it. To assure the signatory of the \texttt{AvailableArbitrators} contract is the real \textit{Operator} party, we confirm that the party is different from the \textit{Tenant} and \textit{Landlord}. Note that, if another party had created the \texttt{AvailableArbitrators}, impersonating the \textit{Operator}, the \texttt{InvokeArbitrators} choice would have failed in the \texttt{MIReport} fetch above. This is because this entity would not be a signatory in the \texttt{LeaseAgreement} contract. For this reason, the \texttt{InvokeArbitrators} choice must be exercised in the \texttt{LeaseAgreement} contract, where the \textit{Operator}, \textit{Tenant}, and \textit{Landlord} are signatories. After these verifications, the \texttt{InviteArbitrators} contract associated with the \texttt{MIReport} is created. 

In the \texttt{InviteArbitrators} contract, all \textit{Arbitrators} can exercise the \texttt{Accept} choice. This adds them to the list of confirmed \textit{Arbitrators}. When the list is full, that is, when its size is equal to the number of arbitrators agreed on in the lease terms. Either the \textit{Tenant} or \textit{Landlord} finalizes the invitation by exercising the \texttt{ConfirmAttribution} choice. This choice gets the confirmed parties from the \texttt{InviteArbitrators} contract to ensure that the \textit{Arbitrators} are set from an \texttt{InviteArbitrators} contract. It then archives the \texttt{InviteArbitrators} and assigns the confirmed \textit{Arbitrators} to the \texttt{MIReport}. 

In the Polling Process, the \textit{Visitor}, will visit the property and create the \texttt{Poll} contract represented below, by exercising the \texttt{CreatePoll} choice in the \texttt{MIReport} contract. This choice can only be exercised by the \textit{Arbitrators} party.

\begin{footnotesize}
\begin{verbatim}
Poll
    tenant : Party
    landlord.: Party
    miDetails : MIdetails
    visitor : Party 
    visitDetails : Text
    assessmentDate : Date
    reparationDate : Date
    cost: Int 
    voters : Parties  
    alreadyVoted : Parties 
    votes : [Responsability]
    
    signatory alreadyVoted, tenant, landlord
    observer  voters
\end{verbatim} 
\end{footnotesize}

This contract is created with the \texttt{alreadyVoted} attribute containing only the \textit{Visitor}. Both the \textit{Tenant} and \textit{Landlord} must be signatory to this contract. The \texttt{voters} attribute contains all \textit{Arbitrators} that will vote (the same that confirmed the invitation).
A vote is submitted by exercising the \texttt{Vote} choice, this archives the contract and creates a new instance with the voter in the \texttt{alreadyVoted} list and its vote in the votes list. The choice verifies that the voter is not already on this list, preventing duplicate votes. The members of this list are signatories of the contract, so they can not deny the vote. When all votes are completed, any entity can finalize the poll by exercising the \texttt{FinalizeVotation} choice. This choice archives the \texttt{Poll} contract and creates a \texttt{MIresultArbitration} contract with the results of the poll, that is, the mean of all votes. This choice asserts the \texttt{voters} set is equal to the \texttt{alreadyVoted} set. In the Arbitrator UI, this choice is executed automatically by the last \textit{Arbitrator} submitting the vote.

\subsection{Back-end Interaction Layer}

This layer consists of the User UI, Arbitrator UI, and the ARC-Oracle API. The interaction with the ledger is enabled by the Daml Java Bindings. The User and \textit{Arbitrator} interact with the system through a command-line interface. The User UI and Arbitrator UI have the same structure: A \texttt{LedgerCommunication} class to interact with the ledger and an \texttt{IO} class to interact with the stdout. The ARC-Oracle API has a \texttt{LedgerCommunication} class. A scheduler could be implemented using the Java \texttt{ScheduledExecutorService} class with a 24-hour period. However, to simplify development and testing, we opted to implement the trigger as input from the stdin.

\subsubsection{Ledger Communication}

In the \texttt{LedgerCommunication} class the necessary parameters to communicate with the ledger are: the ledger's IP address, the port, and the party Ids. 
It has an in-memory contract store that provides a live view of all active contracts it interacts with that maps contract Ids to external Ids. The contract store is updated to a recent state using the Active Contract Service from the Daml Java Bindings library. The \texttt{LedgerCommunication} class has a probe function using the Transaction Service that continuously updates the contract store in response to all ledger events that create or archive contracts. It is important to remember that contracts can only be seen, therefore, stored, when the party has a stake in the contract. The probe function creates a new thread to update the contract store in the background. This function uses the \texttt{Flowable} class of the RxJava library to deal with the backpressure of the events received from the ledger. 
A \texttt{ConcurrentHashMap} is used to map the external Ids to each Contract and a \texttt{HashBiMap} wrapped around a thread-safe BiMap. For the same reason, it is used \texttt{AtomicLong} to get and update the external Ids. 
The contracts stored in each interface are as follows:

\subsubsection*{User UI} 

\begin{inlinelist} 
    \item \textbf{LeaseAgreement:} Live view.
    \item \textbf{MIReport:} Live view.
    \item \textbf{InviteArbitrators:} Not stored, probed to check if agreed number of arbitrators reached.
    \item \textbf{AvailableArbitratorsRequest:} Not stored, created by the user, argument of \texttt{AddObserver}.
    \item \textbf{AvailableArbitrators:} Live view, only one contract in the ledger so stored in most recent contract instance (fetched with public party).
    \item \textbf{IOU:} Not stored, probed when a new instance is created,  notification displayed. 
\end{inlinelist}

\subsubsection*{Arbitrator UI}

\begin{inlinelist} 
    \item \textbf{InviteArbitrators:} Live view but only incomplete invitations where the party has not yet confirmed are stored.
    \item \textbf{MIReport:} Live view.
    \item \textbf{Poll:} Live view.
\end{inlinelist}

\subsubsection*{ARC-Oracle API}

\begin{inlinelist} 
    \item \textbf{DateClock:} Live view, only one contract visible per party in the ledger, so only most recent instance stored. 
    \item \textbf{Evolve:} Same as \textbf{DateClock}. 
    \item \textbf{DateClockUpdateEvent:} No stored, obtained from return value of \texttt{Advance}.
\end{inlinelist}

The commands that are submitted to the ledger are encoded in an instance of the \texttt{Update} class. They are submitted to the ledger by the Command Submission Service from the Ledger API Services. The commands are submitted synchronously. 

\subsubsection{UI Interaction}


In the User UI, there are three navigation menus: Main, Lease Agreements, and MIs menus. The navigation options in each one of the menus are:

\subsubsection*{Main Menu} 

\begin{inlinelist} 
    \item \textbf{Lease Agreements List:} Displays the lease agreements the user is associated with. 
    \item \textbf{Make Proposal:} For the user to create a proposal.
    \item \textbf{Proposals List:} Lists the proposals waiting confirmation.
    \item \textbf{Display all contracts:} Displays the in-memory contract store (for debugging).
\end{inlinelist}

\subsubsection*{Lease Agreements Menu}

\begin{inlinelist} 
    \item \textbf{MIs List:} Lists the MIs associated with the selected lease agreement. 
    \item \textbf{Create Maintenance Issue:} For the user to create a new MI associated with the selected lease agreement.
    \item \textbf{Back to main menu:} Ditto.
\end{inlinelist}

\subsubsection*{Maintenance Issues Menu}

\begin{inlinelist} 
    \item \textbf{Mediation Resolution:} For the user to submit an assessment to resolve the issue.
    \item \textbf{Mediation Proposals:} Lists the mediation proposals for this MI.
    \item \textbf{Arbitration Resolution:} Triggers the Arbitration resolution process by calling the \texttt{invokeArbitrators} method in the \texttt{LedgerCommunication instance}.
\end{inlinelist}

\vspace{0.2cm}

In the Arbitrator UI, there is one Main Menu with three navigation options, these are:
\begin{inlinelist} 
    \item \textbf{List invitations to MIs Reports:} Lists all invitations; the \textit{Arbitrator} can then select an invitation and accept or reject.
    \item \textbf{List Assigned MIs Reports:} Lists all MIReport contracts that the \textit{Arbitrator} is associated with. The \textit{Arbitrator} can then select the MI to create a poll. Then he provides the details of the visit, the cost of the repair, the date of the assessment, the date of the repair, and its vote.
    \item \textbf{List Polls:} Lists all polls to which the \textit{Arbitrator} is assigned. He can then select the poll for which he wants to vote. Displays the information provided by the \textit{Visitor}. Based on this info, the \textit{Arbitrator} submits his vote.
\end{inlinelist}

The Arbitrator UI also displays notifications for various events, such as the creation of a new MI invite, the creation of a new poll, or the calculation of poll results.

\section{Evaluation}

The evaluation criteria are different for each oracle. The latency of the ARC-Oracle will be assessed and the MI-Oracle will be compared with the current approach to solving disputes between tenants and landlords. The whole platform will be assessed in terms of security, mainly integrity and privacy.

\subsection{ARC-Oracle Evaluation}

\begin{figure}[t]
  \centering
  \includegraphics[width=0.7\linewidth]{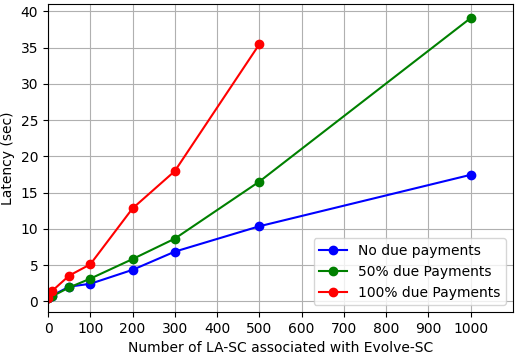}
 \caption{Latency analysis of the ARC-Oracle}
  \label{fig:arc-latency}
\end{figure}

The ARC-Oracle is evaluated against the latency of completing its two transactions, advancing the \texttt{DateClock} contract and verifying and processing the due payments. 
The latency value is calculated as a function of the number of \texttt{LeaseAgreement} contracts associated with the \texttt{Evolve} contract. It was tested with a single ARC-Oracle node running in a Canton ledger with a single node and a single synchronization domain,
in a MacBook with a 1,4 GHz Quad-Core Intel Core i5 processor and 16GB of RAM. Latency was measured with no due payments, half due payments and all due payments. The values are represented in Fig.\ \ref{fig:arc-latency}.  

From the analysis, the number of due payments in the lease agreement smart contracts associated with the Evolve smart contract has an impact on latency. According to what is expected, a higher number of due payments increases the latency of the request. Since rent payments are typically scheduled on the 25th of each month, the red and blue lines provide more descriptive results. Optimizing just for these peak times, the 25th of each month, might allow better resource usage.
There is no point in evaluating the throughput as there is always one transaction per day.

\subsection{MI-Oracle Evaluation}

The MI-Oracle is designed to be an alternative to the traditional litigation process for resolving disputes between tenants and landlords. To evaluate this process, we compare the resources consumed in the current process with those consumed when using the MI-Oracle. 

To compare the duration of a dispute with the traditional litigation process and our solution, we provide an overview of the traditional litigation process. This process begins with a civil action, the legal process through which an entity initiates a lawsuit against another party to resolve a civil dispute. A civil execution follows a civil action if the judgment or order of the civil action is not voluntarily complied with by the losing party. In Lisbon, the average duration of a civil action in 2022 was 12 months and of a civil execution was 56 months \cite{justicaEstatisticas}.  These statistics do not include the time to hire a lawyer. 

Our solution does not include a process to solve a civil execution, as this would require granting the \textit{Arbitrators} judicial power, so the reference used is the average duration of a civil action, 12 months. In Lisbon, there are around 120000 rental agreements per year \cite{cartaMunicipal2023}. 
Using the MI-Oracle process during the lease period reduces the waiting time for the problem solution from months to days. The reason is that the duration depends on the availability of the \textit{Arbitrator} to visit the property, and the \textit{Tenant} and \textit{Landlord}. The duration of traditional litigation depends on factors such as the time it takes to find a lawyer, the availability of both parties, their lawyers and a jury, the court's schedule, among others.

The traditional process requires a lawyer for consultation and representation of each party, as well as a jury. In the MI-Oracle process, it is necessary to allocate a \textit{Visitor} plus the defined number of \textit{Arbitrators} in the lease agreement. If the agreed number of \textit{Arbitrators} is small, the traditional litigation process will use more human resources than our solution. 


\subsection{Platform Security}

The platform is created to provide a proposal service and a lease agreement management service. This security analysis considers a threat any breach in confidentiality or unauthorized access that compromises the protections enforced by the smart contract implementation. We assume that a user party, \textit{Artbitrator} party, and \textit{Operator} party cannot be impersonated, which means that any action attributed to the party was carried out by the individual associated with that party. 

We used the Daml testing tools, mainly Daml Script, for this evaluation. Daml Script is a method for testing Daml models and receiving feedback in Daml Studio, which is an IDE designed specifically to work with Daml. Our platform is in agreement with the security requirements in the creation of the following contracts: \texttt{LeaseAgreement}, \texttt{MIReport}, \texttt{MIReportMediation}, \texttt{Poll}, \texttt{MIReportResults}.
As an example, we show that a lease agreement cannot be created without the consent of both the \textit{Tenant} and \textit{Landlord}. The \texttt{LeaseAgreement} contract cannot be created directly from its template without the consent of the \textit{Tenant}, \textit{Landlord} and \textit{Operator}, because they are all signators. The only alternative way to create a \texttt{LeaseAgreement} contract is for the \textit{Operator} to exercise the \texttt{Approve} choice in the \texttt{LACreationRequest} contract, in this contract both the \textit{Tenant} and \textit{Landlord} are signatories. There is only one way to create this contract, which is for the \textit{Landlord} to exercise the \textit{Accept} choice in the \texttt{Proposal} contract created by the \textit{Tenant}. Before accepting, the \textit{Landlord} can confirm the lease terms, the house, and the \textit{Operator} party. To evaluate the privacy requirements of the lease agreement, the contract's creation was simulated using Daml Script. 

\begin{figure}[t]
  \centering
  \includegraphics[width=0.75\linewidth]{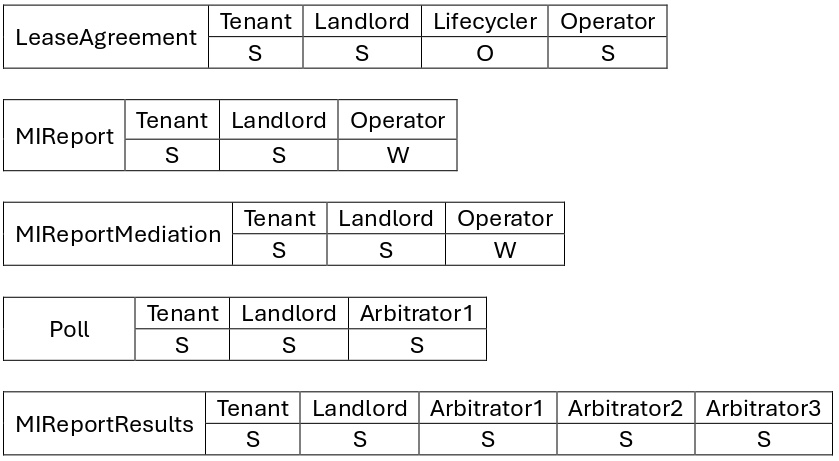}
 \caption{Contract creation results}
  \label{fig:securityResults}
\end{figure}

Fig.\ \ref{fig:securityResults} displays the results generated by the script responsible for creating the contracts described earlier. The \textit{Arbitrator1} is the \textit{Visitor}. The parties who can see the contract are shown on the right side of the boxes. The signatories are represented with an S and the observers are represented with an O. The W in the indicates that the party witnessed the creation of the contract. In the MIReport, for example, it is because the \textit{Operator} is a signatory in the \texttt{LeaseAgreement} contract and the \texttt{MIReport} contract creation is a consequence of an action in the \texttt{LeaseAgreement} contract. This means that the \textit{Operator} can read the contract but cannot act on it or see the consequences of actions on it. 

\section{Conclusion}

This paper introduced two oracles for blockchain-supported property rental, one for automating payments and another for supporting MI management. 
The paper also presents a property rental platform that is used to assess the two oracles. 
This work focused first on finding the time-based trigger solution that best aligned with the needs of the rental platform. 
It relies on obtaining the transaction timestamp and converting it into a date value. 
%
The disputes in a property rental environment have characteristics that make it a more challenging problem to address: the number of issues that can appear in a property, the property being a private and static structure, and the subjective aspect of the assessment. These characteristics require us to use humans to perform the assessment, implying an oracle with UI. 

\bibliographystyle{IEEEtran}
\bibliography{sample-base}

\end{document}